\documentclass{mem}
\usepackage[authoryear]{natbib}
\usepackage{txfonts}
\usepackage{balance}
\usepackage{graphicx}
\usepackage[a4paper]{hyperref}
\idline{75}{1}

\begin{document}
\def\i{{\,\scriptsize I}}
\def\ii{\,{\scriptsize II}}

\title{
Radio Spectroscopy of Active Galactic Nuclei
}

   \subtitle{}

\author{
A.~P. Lobanov
          }

  \offprints{A.~P. Lobanov}

\institute{
Max-Planck-Institut f\"ur Radioastronomie, Auf dem H\"ugel 69,
53121 Bonn, Germany \\
\email{alobanov@mpifr-bonn.mpg.de}
}

\authorrunning{Lobanov}

\titlerunning{Radio Spectroscopy of AGN}

\abstract{Radio spectroscopy offers a number of tools for studying a
large variety of astrophysical phenomena, ranging from stars and their
environment to interstellar and intergalactic medium, active galactic
nuclei (AGN) and distant quasars. Main targets of extragalactic radio
spectroscopy are atomic and molecular material in galaxies, H\ii\ regions,
and maser emission originating in the dense, circumnuclear
regions. These studies cover all galactic types and span an impressive
range of angular scales and distances.  Molecular emission, hydrogen
absorption and maser lines have become the tools of choice for making
an assessment of physical conditions in the nuclear regions of
galaxies.  In this contribution, some of the recent advances in the
aforementioned fields will be reviewed and discussed in connection
with future radio astronomical facilities.

}
\maketitle{}

\section{Introduction}

\begin{figure}
\begin{center}
\includegraphics[width=0.37\textwidth,angle=-90,bb=0 0 516 711,clip=true]{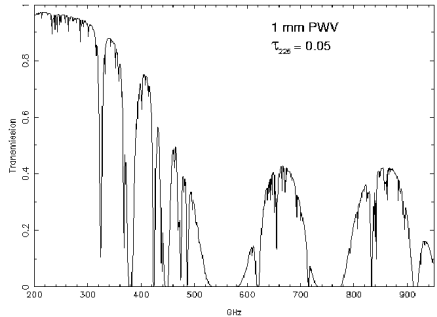}
\caption{Atmospheric transmission for frequencies up to 1000\,GHz.
Below 500\,GHz ($\lambda =0.6$\,mm), absorption due to water in the
Earth atmosphere does not play a significant role and a large number
of atomic and molecular lines can be observed using ground based
facilities working in centimeter and millimeter bands. Above 500\,GHz
and outside the two main windows at 350\,$\mu$m and 450\,$\mu$m,
observations can only be made with airborne and space observatories.
(Courtesy of the Steward Observatory Radio Astronomy Laboratory.) 
\label{lobanov-fig1}}
\end{center}
\end{figure}

Spectroscopic observations in the radio domain present remarkable
opportunities to study chemical composition and kinematics of plasma,
gas, and molecular material in stars, galaxies and galactic nuclei
\citep{verschuur1988}. This makes radio spectroscopy an excellent
counterpart to spectral studies made in other bands. Radio
spectroscopy spans an impressive spectral range of frequencies
(0.1\,GHz to 500\,GHz, Fig.~\ref{lobanov-fig1}), reaches out to redshifts in
excess of 5 \citep{klamer2005}, and operates at angular scales down to
a fraction of milliarcsecond, using the techniques of connected and
long baseline interferometry. Radio spectroscopic studies of active
galaxies (AG) and active galactic nuclei (AGN) embrace both broad-band
and narrow band spectral measurements, probing non-thermal emission
from cosmic plasmas as well as thermal and maser radiation from
various atomic and molecular species.  Observations of water masers in
AGN have paved the way to detailed studies of inner accretion disks
and accurate measurements of black hole masses \citep{moran1995}.
Measurements of synchrotron self-absorption have been used to
determine the physical conditions in AGN on sub-parsec scales
\citep{lobanov1998a}.  Hydrogen and free-free absorption toward bright
continuum sources has become a powerful tool to probe the
circumnuclear environment in galaxies \citep{conway1999,walker2000}.
Physical conditions in relativistic outflows from AGN can be probed
effectively by studying the distribution and evolution of the turnover
frequency of the synchrotron emission from parsec-scale jets
\citep{lobanov1998b}.  Observations of molecular lines, the CO lines
in particular, have become a tool of choice to probe the galactic
activity up to very high redshifts \citep{klamer2005}.  Radio
recombination lines \citep{kardashev1959} have been detected so far
only in a small number of extragalactic objects \citep{gordon2002} and
can be used as an effective indicator of the physical state of ionized
material in galaxies. The next generation of radio astronomical
instruments: LOFAR (at 10--240\,MHz), SKA (300\,MHz--43\,GHz), and
ALMA (30--1000\,GHz) will be most actively involved in radio
spectroscopy, achieving up to two orders of magnitude increase in
sensitivity and opening new horizons for spectroscopic studies in the
radio domain.
 
\section{Atomic lines}

With the exception of recombination lines, atomic radio lines are
rare, as most atomic transitions produce spectral lines at wavelengths
in the infrared or shorter. Table~\ref{lobanov-tb1} lists the atomic
radio lines, including the famous 21\,cm (1.42\,GHz) hyperfine line of
neutral hydrogen. The D\i\ line at 327\,MHz is the deuterium analogue
of the 21\,cm line of H\i. The D\i\ line has not yet bet detected,
even inside our Galaxy \citep{heiles1993}, probably because most of
the deuterium in the Universe is in the molecule HD
\citep{tielens1983}. The radio lines of both H\i\ and D\i\ are
cosmologically important, as they can be used to measure directly the
primordial composition of matter in the Universe. The hyperfine
spin-flip transition of $^{3}$He$^+$ has been detected in a number of
Galactic sources \citep{bania1997}. The C\i\ lines are used as tracers
of molecular gas in the central regions of galaxies \citep{israel2002}
and at high redshifts \citep{weiss2003,papadopoulos2004}. 

\begin{table}[h]
\caption{Parameters of atomic radio lines}
\label{lobanov-tb1}
\begin{tabular}{llrr}\hline\hline
Line & Transition & $\nu$/GHz & Det. \\ \hline
D\i    & $^{2}S_{1/2}$, $F=3/2-1/2$ & 0.327 & no \\
H\i   & $^{2}S_{1/2}$, $F=1-0$     & 1.420 & G,E\\
$^{3}$He$^+$ & $^{2}S_{1/2}$, $F=0-1$ & 8.665 & G\\
C\i   & $^3P_1-$$^3P_0$ & 492.16 & G,E \\
C\i   & $^3P_2-$$^3P_1$ & 809.34 & G\\ \hline
\end{tabular}
{\bf Note:}~The last column shows whether a given line has been detected in Galactic (G) and extragalactic (E) objects.
\end{table}

\subsection{Hydrogen line}

The 21\,cm line of H\i\ is a ``work horse'' of spectroscopy in the
radio domain.  It has been used most successfully to study the
characteristics of diffuse interstellar medium in the Milky Way and in
other galaxies.  The H\i\ line is characterized by its spin
temperature $T_\mathrm{s}$ (the excitation temperature for hyperfine
transitions). Several processes contribute to $T_\mathrm{s}$,
including (i)~absorption and emission stimulated by external photon
radiation field; (ii)~collisions with hydrogen atoms and free
electrons; (iii)~``pumping'' by Ly\,$\alpha$ photons \cite{field1959}.
This allows estimates of various properties of H\i\ emitting material
to be made \citep{rohlfs2004}.  In addition, the Zeeman effect of a
frequency shift of the circularly polarized components of H\i\ line
profile (both in emission and in absorption) enables direct
measurements of the line-of-sight magnitude of the magnetic field.

Observations of neutral hydrogen in galaxies is a prime tool for
studying the distribution and motion of interstellar gas in galaxies.
For a galaxy
at a distance $d$, the mass of neutral hydrogen can be estimated from
\[
M_\mathrm{HI}[M_{\odot}] = 2.36 \times 10^5\,d^2 \int S_\nu \mathrm{d}\,v\,,
\]
where $S_\nu$ is in Jy and the line is integrated over velocity in
km\,s$^{-1}$.  Observations of H\i\ line have been used to study
galactic rotation and determine the mass distribution in a number of
spiral galaxies \citep{bosma1981}. These have provided the basis for
the virial mass estimates for the central regions of galaxies and for
the famous Tully-Fisher (\citeyear{tully1976}) relation connecting the
maximum rotational velocity and luminosity of galaxies. Mapping the
H\i\ distribution in relatively close galaxies shows that (i)~the H\i\ 
is not centrally concentrated and its extent is larger than the
optical extent; (ii)~the rotation appears to be flattening at large
distances (in contrast to the decrease observed in the optical
rotation curve); (iii)~H\i\ links and bridges exist between galaxies
separated by a few tens of kiloparsecs. At present, the brightness
sensitivity of existing instruments sets an angular resolution limit
of $\sim 5$\,arcsec for studies of H\i\ emission and precludes
detailed measurements close to active nuclei. The situation should
change with the SKA providing a factor of 100 improvement of
sensitivity. It should also be noted that, besides galactic studies,
observations of H\i\ have a paramount importance for cosmological
investigations, most notably for studies of structure formation and
the epoch of re-ionization.

\subsection{Recombination lines}

\begin{figure}
\begin{center}
\includegraphics[width=0.45\textwidth,angle=0,bb=12 200 582 629,clip=true]{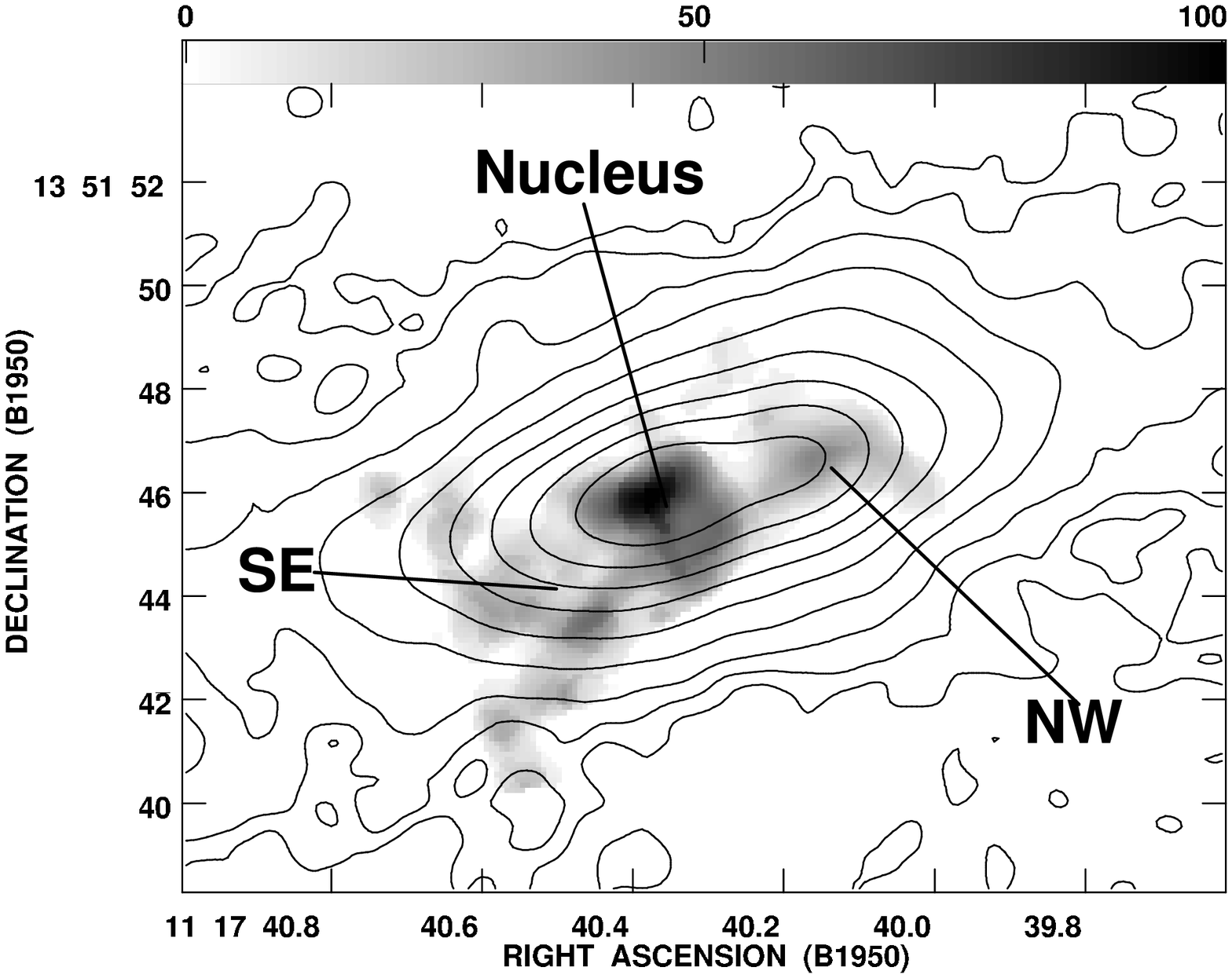}
\caption{Integrated H92$\alpha$ line flux image of the nuclear region
  of NGC\,3628. The line is produced by the $\alpha$-transition from
  the energy level with $n=92$ ($\lambda = 3.608$\,cm, ). The
  gray-scale range is 0--100 mJy\,beam$^{-1}$\,km\,s$^{-1}$. The
  contours represent the continuum flux densities of 0.04, 0.08, 0.16,
  0.32, .  . . , 10.3 mJy\,beam$^{-1}$. Labeled are three RRL emission
  components: Nucleus, SE, and NW. The position-velocity distribution
  of the RRL emission implies a rotating circumstellar disk with a
  total dynamical mass of $3\times 10^8\,\mathrm{M}_\odot$ within a
  distance of 120\,pc from the nucleus \citep{zhao1997}.
\label{lobanov-fig2}}
\end{center}
\end{figure}

Recombination lines are produced by electrons bound to ions and
cascading downwards in the energy levels. Observability of
recombination lines in the radio regime was predicted by Kardashev
(\citeyear{kardashev1959}). There is a large number of radio
recombination lines (RRL) in the 1--500\,GHz range. Wavelengths of
recombination lines of an atom with a total mass $M$ are given by
the well-known Rydberg formula 
\[
\lambda(n,\Delta\,n) \simeq \frac{1}{R_\mathrm{M}\, Z^2}\left[ \frac{1}{n^2} -
\frac{1}{(n+\Delta n)^2}\right]^{-1}\,,
\]
where $R_\mathrm{M}$ is the atomic Rydberg constant, $Z$ is the
effective charge of the nucleus, $n$ is the lower principal quantum
number, and $\Delta n \ll n$ is the change in $n$.  Thus, for
$\alpha$-transitions, with $\Delta n = 1$, the wavelengths of
recombination lines exceed 1\,cm for $n>60$ (see
Fig~\ref{lobanov-fig2}).

RRL can be used to
estimate temperatures (from the ratio of line energy to that of the
underlying free-free continuum), densities (from the line broadening), and
kinematics (from the Doppler modification of the line shape and center
frequency).  
RRL have been observed throughout our own Galaxy
\citep{gordon2002}, most notably in the Galactic H\ii\ regions
\citep{georgelin1976,odegard1985}. The first extragalactic detection
of a recombination line was made almost 30 years ago in M\,82
\citep{shaver1977}, but the total number of RRL detections in other
galaxies is small, and no detections of RRL in quasars have been
reported so far.  Observations of RRL over a wide range of quantum
levels provides valuable information on the physical state of the
ionized gas \citep{anantha1993}. High-resolution observations of RRL
(Fig.~\ref{lobanov-fig2}) can be used determine velocity fields in
galactic nuclei at an arcsecond resolution, and dynamics of the galaxies
and especially the nuclear regions can be studied \citep{zhao1997}.

\section{Molecular lines}

Molecular line emission is produced via three different type of energy
transitions: (i)~electronic transitions, with typical energies of $\sim
5$\,eV (gives lines in the visual and UV spectrum); (ii)~vibrational
transitions (from oscillations of the relative positions of nuclei
with respect to their equilibrium positions), with typical energies of
0.01--0.1\,eV corresponding to lines in the infrared band;
(iii)~rotational transitions, with typical energies of $\simeq
10^{-3}$\,eV corresponding to lines in the millimeter and centimeter
ranges. 

\begin{figure}
\begin{center}
\includegraphics[width=0.45\textwidth,angle=0,bb=0 70 582 757,clip=true]{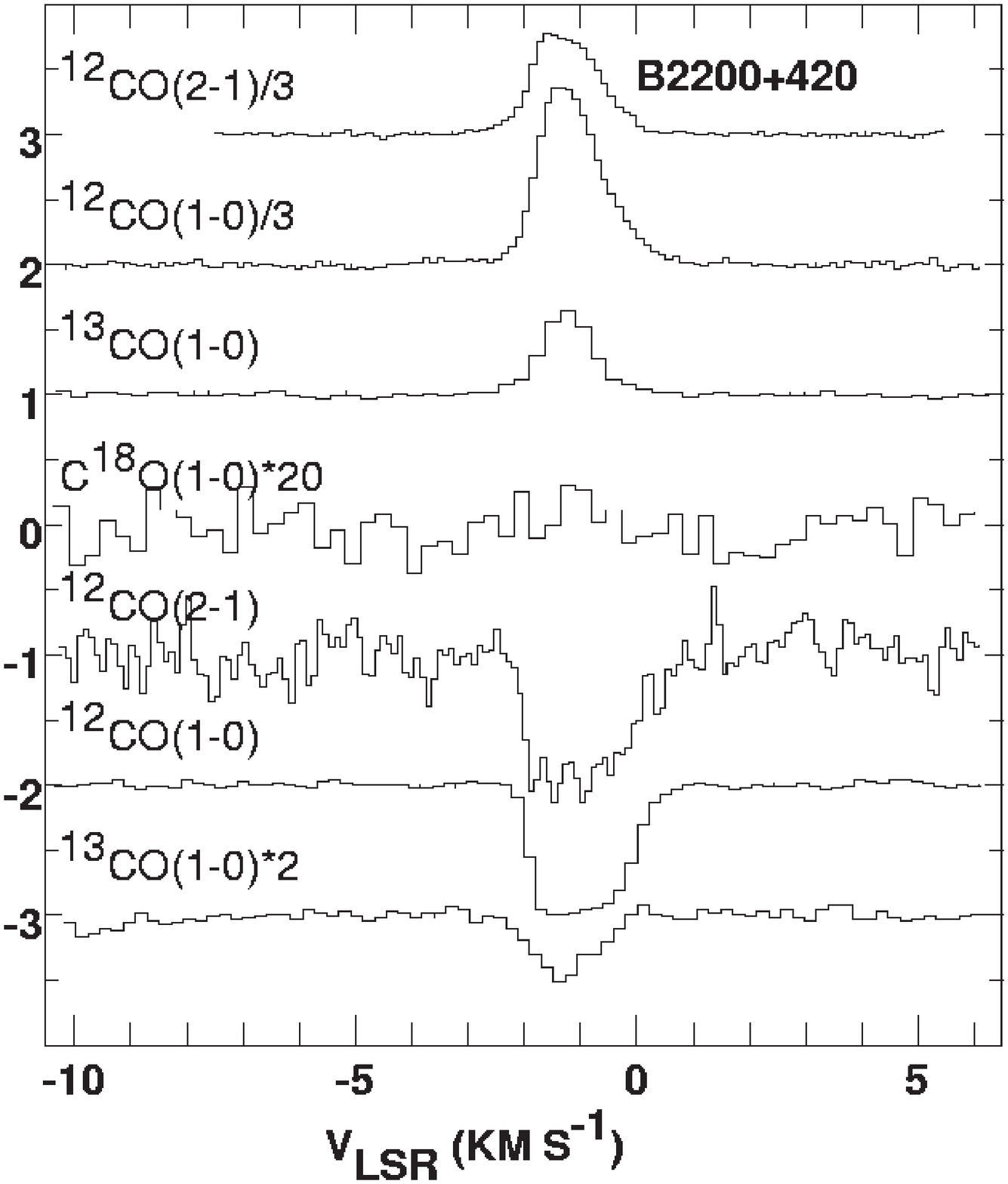}
\caption{CO absorption and emission profiles toward BL\,Lac. The
spectra indicate a surprisingly one-dimensional structure of the
molecular gas, as a fully-formed line profile is observed from an
integration along the line of sight to a point source of $\sim 1$\,mas
in size. This implies that the material producing the molecular
emission is highly structured \citep{liszt1998}.}
\label{lobanov-fig3}
\end{center}
\end{figure}

Molecular line radio astronomy began in 1963, with a detection of
absorption by the OH molecules of continuum radio emission from
supernova remnant Cassiopeia\,A. Contrary to the atomic, radio lines
from molecules provide a vastly richer field of study, with a total of
over 120 molecular species identified in the circumstellar medium and
more than 20 molecular species detected toward extragalactic regions
\citep{mauersberger1993,mauersberger2003}. Molecular gas has been
detected in more than 650 galaxies \citep{verter1990}, and the number
of detections is increasing every year. Among the most common
molecules detected in other galaxies (Table~\ref{lobanov-tb2}) are
hydroxyl (OH), methanol (CH$_3$OH), formaldehyde (H$_2$CO), ammonia
N\,H$_3$, hydrogen cyanide (HCN), formylium (HCO$^+$), silicon
monoxide (SiO), and carbon monoxide (C\,O).  Ammonia (NH$_3$) is a good
temperature indicator of extragalactic molecular gas.  The CO and
HCN emission and absorption are good tracers of the mass of the
molecular gas in galaxies, and they can nowadays be detected over an
enormous range of column density at millimeter wavelengths \citep{liszt1998}.
Observations of CO toward powerful AGN imply that the molecular gas is
highly structured in the nuclear regions (Fig.~\ref{lobanov-fig3}).
However, here again, the sensitivity of modern instruments does not permit
direct studies of molecular emission at a resolution needed to probe
directly the conditions in circumnuclear regions in galaxies.

\begin{table}[h]
\caption{Commonly observed molecular lines}
\label{lobanov-tb2}
\begin{tabular}{llr}\hline\hline
Line & Transition & $\nu$/GHz \\ \hline
OH & $^2\Pi_{3/2}F = 1 - 2$ & 1.61223 \\
OH$^\dag$ & $^2\Pi_{3/2}F = 1 - 1$ & 1.66540 \\
OH$^\dag$ & $^2\Pi_{3/2}F = 2 - 2$ & 1.66736 \\
OH & $^2\Pi_{3/2}F = 2 - 1$ & 1.720529 \\
CH$_3$OH$^\dag$ & $J_K = 5_1 - 6_0A^+$ & 6.66852 \\
CH$_3$OH$^\dag$ & $J_K = 2_0 - 3_1E$ & 12.17859 \\
H$_2$CO & $J_{K_{\alpha}K_{c}} = 2_{11}-2_{12}$ & 14.48849 \\
H$_2$O$^\ddag$ & $J_{K_{\alpha}K_{c}} = 6_{16}-5_{23}$ & 22.23525 \\
NH$_3$ & $(J,K) = (1,1) - (1,1)$ & 23.69451 \\
NH$_3$ & $(J,K) = (2,2) - (2,2)$ & 23.72263 \\
SiO$^\dag$    & $J=1-0,v=2$ & 42.82059 \\
SiO$^\dag$    & $J=1-0,v=1$ & 43.12208 \\
SiO    & $J=1-0,v=0$ & 43.42386 \\
SiO$^\dag$    & $J=2-1,v=2$ & 85.64046 \\
SiO$^\dag$    & $J=2-1,v=1$ & 86.24344 \\
SiO    & $J=2-1,v=0$ & 86.84700 \\
HCN$^\dag$    & $J=1-0,F=2-1$ & 88.63185 \\
HCO$^+$& $J=1-0$ & 89.18852 \\
C$^{18}$O & $J=1-0$ & 109.78218\\
$^{13}$CO & $J=1-0$ & 110.20137\\
CO & $J=1-0$ & 115.27120\\\hline
\end{tabular}
{\bf Notes:}~$\dag$ -- often found in maser transitions; $\ddag$ -- always found in maser transitions (from \citealt{rohlfs2004})
\end{table}

\section{Maser lines}

The first detection of an astronomical maser was made thirty years ago
\citep{weaver1965}, with a definitive answer about the maser nature of
the object coming from very long baseline (VLBI) observations
\citep{moran1968}. Subsequently, a number of Galactic and
extragalactic masers have been discovered and studied. Five major
types of cosmic masers are known to date (Table~\ref{lobanov-tb2}):
(i)~H$_2$O masers at 22.2\,GHz (1.3\,cm); (ii)~OH masers at 1.67\,GHz
(18\,cm); (iii)~methanol (CH$_3$OH) masers at 6.67\,GHz (5\,cm);
(iv)~silicon monoxide (SiO) masers at 43.1\,GHz (0.7\,cm) and 86.2\,GHz
(0.3\,cm); (v)~HCN masers at 88.6\,GHz (0.3\,cm) --- and many other
molecules exhibit weaker maser emission. All Galactic masers are
produced in the circumstellar envelopes and near young stars.

\subsection{OH megamasers}

\begin{figure}
\begin{center}
\includegraphics[width=0.45\textwidth,angle=0,bb=0 0 551 576,clip=true]{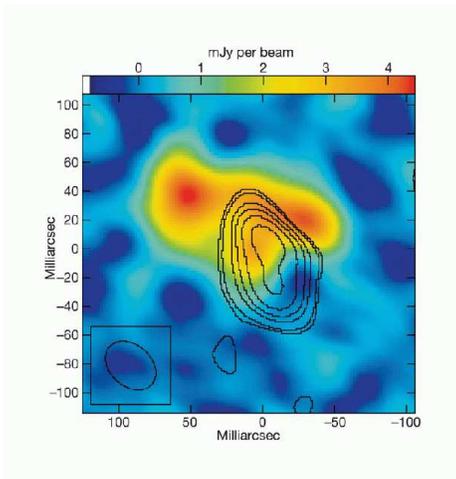}
\caption{The integrated OH (pseudo-color) and radio-continuum
(contours) emission in the nucleus of Mrk\,231. The hydroxyl
(mega)-maser emission shows the characteristics of a rotating, dusty,
molecular torus (or thick disk) located within 30--100\,pc from the
central engine \citep{kloeckner2003}.}
\label{lobanov-fig4}
\end{center}
\end{figure}

Extragalactic maser emission was first discovered in NGC 253 in the
1665\,MHz and 1667\,MHz transitions of OH \citep{whiteoak1974} with a
luminosity of about 100 times greater than typical Galactic OH
masers. In 1982 an OH ``megamaser'' was discovered in Arp220 (IC 4553)
\citep{baan1982}. Since then, more than 50 extragalactic OH sources
have been discovered \citep{baan1993}. The OH megamasers are
characterized by much higher luminosities and broader emission lines
(up to several 100\,km\,s$^{-1}$) compared to Galactic OH masers.  The
standard model of the OH megamasers assumes a low-gain amplification in a
molecular disk around the nucleus of the galaxy
\citep{norris1984}. The gas is pumped by far-infrared radiation
\citep{baan1989} and amplifies the nuclear continuum emission.
Extragalactic OH megamasers often found in ultra-luminous FIR galaxies
(ULIRG), which are generally characterized by morphological
peculiarities, interactions with neighbors, highly ionized species
showing starburst and AGN-type nuclear phenomena
\citep{kloeckner2005}. The OH megamasers are reported to have a
two-component structure (Fig.~\ref{lobanov-fig4}), with a compact and
diffuse maser emission, possibly tracing an interaction between the
ionization cones of the outflow and the molecular torus
\citep{kloeckner2003}.

\subsection{H$_2$O megamasers}

In 1977, a small number of H$_2$O masers have been detected in several
nearby spiral galaxies \citep{churchwell1977}, all of them similar to
the Galactic masers. Extragalactic ``megamasers'', with luminosities
in excess of $100\,L_\odot$, were discovered in nuclei of active
galaxies \citep{dossantos1979,claussen1986}, and soon after became an
impressive tool to study the circumnuclear regions in AGN (with more
than 20 megamasers known to date; \citealt{maloney2002}). Most of the
H$_2$O megamasers have been detected in Seyfert\,2
and LINER galaxies, but the situation has changed with the detection
of a maser in the radio-loud galaxy 3C\,403 \citep{tarchi2003}.
 
All known H$_2$O megamasers have the following common properties: (i)~host
galaxies show evidence for nuclear activity (Braatz et
al. \citeyear{braatz1996,braatz1997}); (ii)~the maser emission is
centered on the nucleus; (iii)~the spatial diameter of the emitting
regions are small ($<3$\,pc) \citep{greenhill1995,miyoshi1995}.

\begin{figure}
\begin{center}
\includegraphics[width=0.45\textwidth,angle=0,bb=49 185 559 571,clip=true]{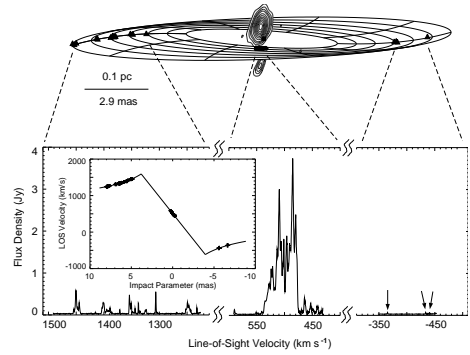}
\caption{H$_2$O masers in NGC\,4258. Top panel: actual maser positions
  (filled triangles and circles) superimposed on the radio-continuum
  emission (contours) and approximated by a model of warped disk. The
  filled square marks the best-fit location of the center of the disk.
  Bottom panel: total spectrum of the maser emission and line-of-sight
  velocities of individual spots fitted by a Keplerian disk model. The
  high-velocity masers trace a Keplerian curve to better than 1\%.
  Observations of H$_2$O masers in NGC\,4258 have allowed to infer a
  geometric distance of $7.2\pm0.3$\,Mpc to the galaxy from the direct
  measurement of orbital motions in the maser spots. The motions imply
  a central object with a mass of $(3.9\pm0.1)\times
  10^{7}\,\mathrm{M}_{\odot}$ \citep{herrnstein1998}.}
\label{lobanov-fig5}
\end{center}
\end{figure}

Extragalactic H$_2$O masers are used in a variety of studies including
(i)~mapping accretion disks in AGN, (ii)~determining the mass of the
central black holes, (iii)~obtaining geometric distances to
galaxies (Fig.~\ref{lobanov-fig5}; \citealt{herrnstein1998}), (iv)~studying
the interaction between the dense molecular material and the
ionization cones \citep{gallimore1996,greenhill2001} and nuclear jets
\citep{claussen1998,peck2003}, and (v)~measuring proper motions of maser
spots in the local group of galaxies \citep{brunthaler2005}.

\section{Absorption lines}

\begin{figure}
\begin{center}
\includegraphics[width=0.45\textwidth,angle=0,bb=0 0 583 445,clip=true]{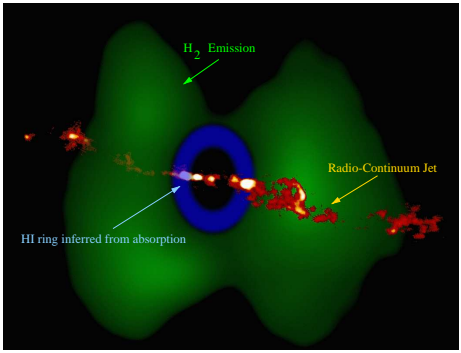}
\caption{Montage of the inner 250 pc of NGC 4151. The H$_2$ emission traces 
a torus, the H\i\ absorption comes from a ring inside the torus. Ionized gas (black) is assumed to fill the torus inside the H\i\ ring
\citep{mundell2003}.}
\label{lobanov-fig6}
\end{center}
\end{figure}

In the radio domain, absorption due to several atomic and molecular
species has been detected, most notably due to H\i, CO, OH, and
HCO$^+$ (see Table~\ref{lobanov-tb2}). In many cases, absorbing
material lies in our own galaxy \citep{liszt2000}. In extragalactic
objects, OH absorption has been used to probe the conditions in warm
neutral gas \citep{goikoechea2004,kloeckner2005}. Observations of CO
and H\i\ absorption have been used to study the molecular tori
\citep{conway1999,pedlar2004} at a linear resolution often smaller
than a parsec \citep{mundell2003}. These observations have revealed,
in particular, the morphology and kinematics of neutral gas in the
molecular torus in NGC\,4151 (Fig.~\ref{lobanov-fig6}) and in a
rotating outflow surrounding the relativistic jet in 1946$+$708
\citep{peck2001}.

\section{Broad-band spectrum}

\begin{figure}
\begin{center}
\includegraphics[width=0.35\textwidth,angle=-90,bb=55 34 578 734,clip=true]{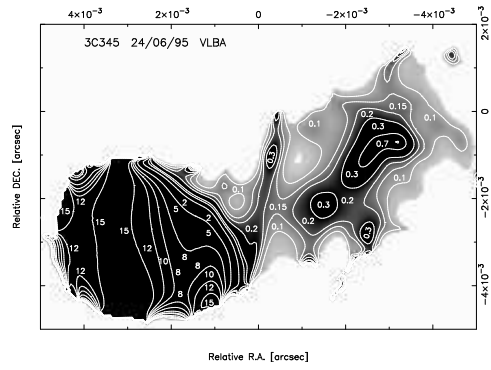}
\caption{Distribution of the turnover frequency in the jet of 3C\,345
obtained from fitting multifrequency VLBI data. Values of the turnover
frequency (in GHz) are shown next to the corresponding
contours. Values below 1\,GHz are upper limits. The distribution
implies the presence of strong shocks within $\sim 1$\,mas ($\sim
5$\,pc) of the nucleus. At larger distances, the shocks dissipate and
give way to Kelvin-Helmholtz instabilities indicated by oblique
patterns of higher turnover frequency  \citep{lobanov1998b}.}
\label{lobanov-fig7}
\end{center}
\end{figure}

Broad-band emission from AGN is dominated by non-thermal synchrotron
and inverse-Compton radiation from highly accelerated plasma. The
synchrotron mechanism plays a more prominent role in the radio domain,
and the properties of the emitting material can be assessed using the
turnover point in the synchrotron spectrum \citep{lobanov1998b},
synchrotron self-absorption \citep{lobanov1998a}, and free-free
absorption in the plasma \citep{walker2000,kadler2004}. The free-free
absorption studies indicate the presence of dense, ionized
circumnuclear material with $T_\mathrm{e} \approx 10^4$\,K distributed
within a fraction of parsec from the central nucleus
\citep{lobanov1998a,walker2000}. Mapping the turnover frequency
distribution (Fig.~\ref{lobanov-fig7}) provides a sensitive
diagnostics of shocks and plasma instabilities in the
relativistic flows \citep{lobanov1997,lobanov1998b}.

\section{Conclusions}

Radio spectroscopy offers a variety of tools that can be used to
assess and quantify the physical conditions in galaxies and active
galactic nuclei. Observations of neutral hydrogen, recombination and
molecular lines provide a wealth of information about the gas dynamics
on galactic scales. However, the limited sensitivity of present-day
instruments does not allow to use the line emission to study
structures on sub-arcsecond scales. 

Maser lines and absorption lines can be studied at highest resolution,
even with the existing instruments. Maser emission, although detected
in a modest number of extragalactic objects, provides an impressive
opportunity to detect and monitor processes in accretion disks on
scales of less than a parsec. Absorption towards compact extragalactic
jets presents an effective tool to diagnose the properties of
circumnuclear gas associated with the molecular tori and
sub-relativistic outflows in AGN.

The field of radio spectroscopy will be revolutionized with the
next-generation radio astronomy facilities planned to be built in the
near future.  A brightness temperature sensitivity as low as
100-10000\,K at angular scales as small as a few milliarcseconds will
become feasible in the millimeter and centimeter domains.  Quasar
molecular absorption lines will be observed in the spectra of many
sources. The optically-obscured molecular tori and the circumnuclear
starbursts of nearby galaxies will be resolved. The presence of
central black holes will be studied kinematically in a large number of
galaxies. This will open a truly new era in the studies of active
galactic nuclei.

\begin{acknowledgements}
I would like to express my sincere gratitude to the conference
organizers for their great support and warm hospitality during the
meeting.
\end{acknowledgements}

\bibliographystyle{aa}

\begin{thebibliography}{99}

\bibitem[Anantharamaiah et al.(1993)]{anantha1993} Anantharamaiah, K.R., et al. 1993, ApJ, 419, 585.

\bibitem[Baan(1989)]{baan1989} Baan W.A. 1989, ApJ, 338, 804

\bibitem[Baan(1993)]{baan1993} Baan W. A., 1993, in Davis R. J., Booth
R. S., eds, Subarcsecond Radio Astronomy. Cambridge Univ. Press,
Cambridge, p. 324

\bibitem[Baan et al.(1982)]{baan1982} Baan W.A., Wood P.A.D. \&
Haschick A.D. 1982, \apj, 260, L49

\bibitem[Bania et al.(1997)]{bania1997} Bania, T.M., et al. 1997,
\apjs, 113, 353

\bibitem[Bosma(1981)]{bosma1981} Bosma, A. 1981, \aj, 86, 1825

\bibitem[Braatz et al.(1996)]{braatz1996} Braatz, J.A., Wilson, A.S. \& Henkel, C. 1996, ApJS, 106, 51

\bibitem[Braatz et al.(1997)]{braatz1997} Braatz, J.A., Wilson, A.S. \& Henkel, C. 1997, ApJS, 110, 321

\bibitem[Brunthaler et al.(2005)]{brunthaler2005} Brunthaler, A., et al. 2005, Science, 307, 1440

\bibitem[Churchwell et al.(1977)]{churchwell1977} Churchwell, E.A., et
al. 1977\, \aap, 54, 969

\bibitem[Claussen \& Lo(1986)]{claussen1986} Claussen, M.J. \& Lo, K.Y. 1986, \apj, 308, 592

\bibitem[Claussen et al.(1998)]{claussen1998} Claussen, M.J., et al. 1998, \apj, 500, L129

\bibitem[Conway(1999)]{conway1999} Conway, J.E. 2005, NewAR, 43, 509

\bibitem[DiMatteo et al.(2002)]{dimatteo2002} DiMatteo, T., et
al. 2002, \apj, 564, 576

\bibitem[dos Santos \& Lepine(1979)]{dossantos1979} dos Santos, P.M. \& Lepine, J.R.D. 1979, Nature, 278, 34

\bibitem[Field (1959)]{field1959} Field, G.B. 1959, \apj, 129, 536

\bibitem[Gallimore et al.(1996)]{gallimore1996} Gallimore, J.F., et al. 1996, \apj, 462, 740

\bibitem[Georgelin \& Georgelin(1976)]{georgelin1976} Georgelin,
Y.M. \& Georgelin, Y.P. 1976, \aap, 49, 57

\bibitem[Goikoechea et al.(2004)]{goikoechea2004} Goikoechea, J.R.,
Mart\'in-Pintado, J. \& Chernicharo, J. 2004, \apj, 619, 291

\bibitem[Gordon \& Sorochenko(2002)]{gordon2002} Gordon, M.A. \&
Sorochenko, R.L. 2002, Radio Recombination Lines, Their Physics and
Astronomical Applications, (Kluwer: Dordrecht)

\bibitem[Greenhill et al.(1995)]{greenhill1995} Greenhill, L.J., et al. 1995, \apj, 304, 21.

\bibitem[Greenhill et al.(2001)]{greenhill2001} Greenhill, L.J., et
al. 2001, in Proceedings of IAU Symposium 205, R.T. Schilizzi (ed.),
334

\bibitem[Heiles et al.(1993)]{heiles1993} Heiles, C., McCullough, P.R. \& 
Glassgold, A.E. 1993, \apjs, 89, 271

\bibitem[Herrnstein et al.(1998)]{herrnstein1998} Herrnstein, J.R., et al. 1998, Nature, 400, 539

\bibitem[Herrnstein et al.(2005)]{herrnstein2005} Herrnstein, J.R., et al. 2005, \apj {\em (in press)} (astro-ph/0504405)


\bibitem[Israel \& Baas(2002)]{israel2002} Israel, F.P. \& Baas, F. 2002, \aa, 323, 82

\bibitem[Kl\"ockner \& Baan(2005)]{kloeckner2005} Kl\"ockner, H.R. \&
Baan, W.A. 2005, Astrophysics and Space Science, 295, 277

\bibitem[Kl\"ockner et al.(2003)]{kloeckner2003} Kl\"ockner, H.R., Baan, W.A. \& Garrett, M.A. 2003, Nature, 421, 821

\bibitem[Kadler et al.(2004)]{kadler2004} Kadler, M., et al. 2004, \aap, 426, 481

\bibitem[Kardashev(1959)]{kardashev1959} Kardashev, N.S. 1959,
Sov. Astron., 3, 813

\bibitem[Klamer et al.(2005)]{klamer2005} Klamer, I.J., Ekers, R.D.,
Sadler, E.M., et al.  2005, AJ, 621, L1

\bibitem[Liszt \& Lucas(1998)]{liszt1998} Liszt, H.S. \& Lucas,
R. 1998, \aap, 339, 561

\bibitem[Liszt \& Lucas(2000)]{liszt2000} Liszt, H.S. \& Lucas,
R. 1998, \aap, 355, 333

\bibitem[Lobanov(1998a)]{lobanov1998a} Lobanov, A.P. 1998a, A\&A, 379, 90

\bibitem[Lobanov(1998b)]{lobanov1998b} Lobanov, A.P. 1998b, A\&ASS, 120, 30.

\bibitem[Lobanov et al.(1997)]{lobanov1997} Lobanov, A.P., Carrara, E. \& Zensus, J.A. 1997, Vistas in Astronomy, 41, 253

\bibitem[Maloney(2002)]{maloney2002} Maloney, P.R. 2002, PASA, 19, 401

\bibitem[Mauersberger \& Henkel(1993)]{mauersberger1993} Mauersberger,
R. \& Henkel, C. 1993, Rev. Mod. Astronomy, 6, 69

\bibitem[Mauersberger et al.(2003)]{mauersberger2003} Mauersberger,
R., et al. 2003, \aap, 403, 561

\bibitem[Miyoshi et al.(1995)]{miyoshi1995} Miyoshi, M. et al. 1995,
Nature, 373, 127

\bibitem[Moran et al.(1968)]{moran1968} Moran, J.M., et al. 1968,
\apjl, 152, L97

\bibitem[Moran et al.(1995)]{moran1995} Moran, J., Greenhill, L.,
Herrnstein J., et al. 1995, PNAS, 92, 25, 11427

\bibitem[Mundell et al.(1999)]{mundell1999} Mundell, C.G., et al. 1999, MNRAS 304, 481.

\bibitem[Mundell et al.(2003)]{mundell2003} Mundell, C.G., et
al. 2003, \apj, 583, 192

\bibitem[Norris(1984)]{norris1984} Norris R.P. 1984, PASA, 5, 514

\bibitem[Odegard(1985)]{odegard1985} Odegard, N. 1985, ApJS, 57, 571

\bibitem[Papadopoulos et al.(2004)]{papadopoulos2004} Papadopoulos,
P.P., Thi, W.-F. \& Viti, S. 2004, \mnras, 351, 147

\bibitem[Peck \& Taylor(2001)]{peck2001} Peck, A.B. \& Taylor,
G.B. 2001, \apj, 554, 147

\bibitem[Peck et al.(2003)]{peck2003} Peck, A.B., et al. 2003, \apj,
590, 149

\bibitem[Pedlar(2004)]{pedlar2004} Pedlar, A. 2004, \apss, 295, 161

\bibitem[Rohlfs \& Wilson(2004)]{rohlfs2004} Rohlfs, K. \& Wilson,
T.L. 2004, Tools of Radio Astronomy, (Springer: Heidelberg)

\bibitem[Shaver et al.(1977)]{shaver1977} Shaver, P.A., Churchwell, E. \& Rots, A.H. 1977, \aap, 55, 435

\bibitem[Tarchi et al.(2003)]{tarchi2003} Tarchi, A., et al. 2003,
  \aap, 407, L33

\bibitem[Tielens(1983)]{tielens1983} Tielens, A.G.G.M. 1983, \aap, 119, 117

\bibitem[Tully \& Fisher(1976)]{tully1976} Tully, R.B. \& Fisher,
J.R. 1976, \aap, 54, 661

\bibitem[Verschuur \& Kellermann(1988)]{verschuur1988} Verschuur,
G.L. \& Kellermann K.I. (eds.) 1988, Galactic and Extragalactic Radio
Astronomy, (Springer-Verlag: Heidelberg)

\bibitem[Verter(1990)]{verter1990} Verter, F. 1990, PASP, 102, 1281

\bibitem[Walker et al.(2000)]{walker2000} Walker, R.C., et al. 2000,
\apj, 530, 233

\bibitem[Weaver et al.(1965)]{weaver1965} Weaver, H., et al. 1965,
Nature, 208, 29

\bibitem[Weiss et al.(2003)]{weiss2003} Weiss, A., et al. 2003, \aap, 409, L41

\bibitem[Whiteoak \& Gardner(1974)]{whiteoak1974} Whiteoak J.B. \& Gardner F.F.,
1974, \apj, 15, L211

\bibitem[Zhao et al.(1997)]{zhao1997} Zhao, J.-H., et al. 1997, ApJ,
482, 186

\end{thebibliography}

\end{document}